\documentclass[onecolumn,preprintnumbers,amsmath,amssymb,notitlepage]{revtex4}
\usepackage{graphicx}
\usepackage{epsfig}
\usepackage{amsmath}
\usepackage{amssymb}
\usepackage{dcolumn}

\begin{document}

\title{Nonlocal Quantization Principle in Quantum Field Theory and Quantum Gravity}

\author{Martin Kober}

\email{MartinKober@t-online.de}

\affiliation{Kettenhofweg 121, 60325 Frankfurt am Main, Germany}

\begin{abstract}
In this paper a nonlocal generalization of field quantization is suggested. This quantization principle
presupposes the assumption that the commutator between a field operator an the operator of the canonical
conjugated variable referring to other space-time points does not vanish as it is postulated in the usual
setting of quantum field theory. Based on this presupposition the corresponding expressions for the field
operators, the eigenstates and the path integral formula are determined. The nonlocal quantization principle
also leads to a generalized propagator. If the dependence of the commutator between operators on different
space-time points on the distance of these points is assumed to be described by a Gaussian function, one
obtains that the propagator is damped by an exponential. This leads to a disappearance of UV divergences.
The transfer of the nonlocal quantization principle to canonical quantum gravity is considered as well.
In this case the commutator has to be assumed to depend also on the gravitational field, since the distance
between two points depends on the metric field.
\end{abstract}

\maketitle

\section{Introduction}

The most important problem of comtemporary fundamental theoretical physics is the unification of quantum theory
and general relativity. A quantum theoretical setting of general relativity as usual quantum field theory
in analogy to the description of the other fundamental interactions is not possible, since such a description
leads to divergencies, which cannot be cured. This fact seems to indicate that the problem of incorporating
general relativity to a quantum theoretical description of all fundamental interactions is directly connected
to the question, whether the basic postulates of quantum field theory in its usual setting can be transferred
to such a more fundamental description of nature. These basic postulates are related to the idea of the basic
structure of space-time. General relativity is a background independent theory, which in contrast to the
usual setting of quantum field theory, on which the other fundamental interactions of the standard model of
particle physics are based, does not presuppose a nondynamical metric structure of space-time. Quantum theory
on the other hand contains a constitutive element of nonlocality, which becomes manifest in the famous
Einstein-Podolsky-Rosen thought experiment for example. The basic postulates of general quantum theory itself,
this means the general Hilbert space formalism in the setting of Dirac without relation to additional field
theoretic elements, does not contain special assumptions about a space-time structure.
Therefore it seems to be possible that a quantum theoretical description of general relativity as well as a
unification of all fundamental interactions appearing in nature could indeed require a decisive modification
of the postulates usually presupposed in quantum field theory. This holds especially for the basic assumption
of local commutativity or microscopic causality, which implies that the commutator or anticommutator
respectively of field operators at two separated space-time points vanishes.

In accordance with the above explanations in the present paper is suggested a generalization of the quantization
principle as it is presupposed in quantum field theory and as it is transferred to canonical quantum gravity.
Usually, the commutator or anticommutator respectively between the field and the corresponding canonical conjugated
momentum is assumed to be proportional to the delta function. In the present paper this quantization principle is
extended in such a way that the delta function is replaced by a general function, which depends on the distance
between the space-time points the corresponding operators refer to. This implies that the commutator between
the field operator and the corresponding operators describing the canonical conjugated quantity on two separated
space-time points does not vanish anymore. Such a quantization principle establishes a nonlocal structure of the
quantum description of a theory, violates the posulate of local commutativity and thus can be called nonlocal
quantization principle. A natural assumption for the function replacing the delta function is a Gaussian function,
which as limiting case merges the delta function. It will be shown that under this assumption one obtains a
propagator, which corresponds to the propagator of usual quantum field theory multiplied with an exponential
damping factor depending on the squared three momentum. This implies that this generalized quantum field theory
contains no ultra violet divergences. The obtained propagator corresponds to the propagator within the coherent
state approach to noncommutative geometry, which has been developed in \cite{Smailagic:2003rp},\cite{Smailagic:2003yb},\cite{Smailagic:2004yy}, extended in
\cite{Huang:2008yy},\cite{Kober:2010um} and transferred to the gravitational field in    
\cite{Kober:2011am}. In \cite{Modesto:2009qc} also appears a propagator of this structure.
It is also possible to formulate this generalized quantization principle within the path integral formalism.
Such a formulation is obtained by replacing the usual inner product between field eigenstates and eigenstates
of the canonical conjugated variables by the generalized one according to the nonlocal quantization principle.
This results in a generalized path integral expression. If the nonlocal quantization principle is considered
as a fundamental description of nature, then one has of course to assume that it is also valid for the
quantization of the gravitational field. Accordingly it has to be postulated a nonlocal generalization of the commutator
between the three metric and its canonical conjugated momentum in quantum geometrodynamics. With respect to the
quantization of the gravitational field, the intricacy arises that the distance between two space-time points
depends on the metric field. Therefore the commutator also depends on the three metric itself. A Generalized
quantization principle in canonical quantum gravity and quantum cosmology has already been considered in
\cite{Kober:2011uj}. Other considerations of a generalized quantization principle with respect to the
variables of quantum cosmology and black holes can be found in \cite{Majumder:2011ad},\cite{Majumder:2011bv}.
The considerations in \cite{Kober:2011uj},\cite{Majumder:2011ad},\cite{Majumder:2011bv} have transferred the
generalized uncertainty principle in quantum mechanics, see \cite{Maggiore:1993kv},\cite{Kempf:1994su},\cite{Hinrichsen:1995mf} for example, which leads to a minimal length, to the variables describing the gravitational field. Another
generalization of the quantization principle in canonical quantum gravity, which is based on quaternions
and has also been transferred to $N=1$ supergravity, has been developed in \cite{Kober:2015bkv}.
However, these generalizations of the quantization principle in canonical quantum gravity in contrast to the
approach of the present paper remain local. Nonlocality has been considered in the context of quantum field theory
\cite{Yukawa:1950eq},\cite{Yukawa:1950er},\cite{Alebastrov:1973vw},\cite{Marnelius:1974rq},\cite{Wreszinski:1975yz},\cite{Dubnickova:1976jb},\cite{Bandyopadhyay:1976sp},\cite{Buchholz:1978fv},\cite{Efimov:1980qp},\cite{Buchholz:1985xs},\cite{Mannheim:1986ad},\cite{Marino:1987tk},\cite{Mironov:1989hm},\cite{Marino:1989ve},\cite{Bernard:1990ys},\cite{Cornish:1991kg},\cite{Cornish:1991ja},\cite{Barci:1995ad},\cite{Solovev:1998gm},\cite{Esposito:1999hp},
\cite{Amorim:1999mr},\cite{Shirokov:2001zh},\cite{Denk:2003jj},\cite{Jain:2003xs},\cite{Denk:2004pk},\cite{Gorbar:2005sx},\cite{Novello:1994ym},\cite{Novello:1994zi},\cite{Wetterich:1997bz},\cite{Dobado:1997yh},\cite{Ahluwalia:1999yc},
general relativity
\cite{Chen:2001gx},\cite{Mukohyama:2001jv},\cite{Bergman:2001rw},\cite{Barvinsky:2003kg},\cite{Hehl:2008eu},\cite{Koshelev:2008ie},\cite{Bronnikov:2009az},\cite{Calcagni:2010ab},\cite{Blome:2010xn},\cite{Chicone:2011me},\cite{Barvinsky:2011rk},\cite{Elizalde:2011su}
and quantum gravity
\cite{Rayner:1989wd},\cite{Cornish:1991iz},\cite{Kim:1993iua},\cite{Barvinsky:1994hw},\cite{Esposito:1998hp},\cite{Shojai:1999jj},\cite{Dominguez:1999sq},\cite{Moffat:2002ce},\cite{Reuter:2002kd},\cite{vonBorzeszkowski:2002ri},\cite{Reuter:2003yb},\cite{Giddings:2006be},\cite{Barvinsky:2011hd}
and also of black holes \cite{Nicolini:2012eu}. But the concept of a nonlocal qusantization principle
has not been studied yet.

The paper is structured as follows:
The generalized formalism of quantum field theory based on this nonlocal quantization principle for a scalar field
is considered at the beginning. Subsequently, the corresponding generalized expression for the propagator is calculated.
Especially the mentioned special case is regarded, where the function the commutator depends on is assumed to be a
Gaussian function leading to a propagator containing an exponential damping function, which depends on the squared
three momentum with negative sign of the exponent as additional factor. This means that the corresponding quantum
field theory contains no ultra violet divergences. After this the nonlocal quantization principle is considered in
the context of canonical quantum gravity and the generalized representations of the operators as well as the
generalized constraints are formulated. At the end the nonlocal quantization principle is transferred to the
path integral formula.

\section{Nonlocal Quantization Principle}

In this section is presented the basic idea of the nonlocal quantization principle in a general way. 
According to the explanations given in the introduction a generalization of the fundamental quantization
principle with respect to the quantization of fields is suggested. This quantization principle omits the
postulate of microscopic causality. The generalized nonlocal quantization principle is obtained by the
following transition:

\begin{equation}
\left[\hat \Phi(x),\hat \Pi(y)\right]=i\delta^3(x-y)\quad \longrightarrow \quad
\left[\hat \Phi(x),\hat \Pi(y)\right]=i\alpha^3(x-y),
\label{generalization_commutator}
\end{equation}
where $\hat \Phi(x)$ denotes a field operator, $\hat \Pi(x)$ denotes the corresponding canonical conjugated
operator, $x$ and $y$ denote three-vectors, $\delta^3(x-y)$ denotes the three-dimensional delta function
referring to a spacelike hypersurface and $\alpha^3(x-y)$ denotes a general function on the spacelike
hypersurface, which is symmetric in $x$ and $y$ and thus just depends on the distance of the points
$x$ and $y$. This assumption is natural, if homogenity and isotropy of space shall be maintained.
The field operator $\hat \Phi(x)$ and the corresponding canonical conjugated operator $\hat \Pi(x)$
are generalizations of the usual operators, which shall be denoted with $\hat \phi(x)$ and $\hat \pi(x)$
and obey the usual commutation relations:

\begin{equation}
\left[\hat \phi(x),\hat \pi(y)\right]=i\delta^3(x-y).
\end{equation}
The generalized operators can be represented in a specific way by using the usual operators.
One possibility is the representation, in which the generalized field operator $\Phi(x)$ is equal
to $\phi(x)$ and the generalized canonical conjugated operator is modified,

\begin{equation}
\hat \Phi(x)=\hat \phi(x),\quad \hat \Pi(x)=\int d^3 z\ \alpha^3(x-z)\ \hat \pi(z),
\end{equation}
and another possibility is the representation, where the generalized canonical conjugated operator
is equal to $\pi(x)$ and the generalized field operator is modified,

\begin{equation}
\hat \Phi(x)=\int d^3 z\ \alpha^3(x-z)\ \hat \phi(z),\quad \hat \Pi(x)=\hat \pi(x).
\end{equation}
The momentum eigenstates in the representation, where the momentum operator is generalized and
where the usual operators are represented with respect to the field, looks as following:

\begin{equation}
|\Pi[\Phi(x)]\rangle=\exp\left[i\int d^3 x\ d^3 y\ {\bar \alpha}^3(x-y)\Phi(x)\Pi(y)\right],
\label{eigenstate_Pi}
\end{equation}
where $\bar \alpha^3(x-y)$ is defined by the following relation:

\begin{equation}
\int d^3 z\ \alpha^3(x-z){\bar \alpha}^3(z-y)=\delta(x-y),
\end{equation}
and $\Pi(x)$ denotes the momentumn eigenvalue corresponding to the momentum eigenstate.
This can be seen as follows:

\begin{eqnarray}
\hat \Pi(x)|\Pi\rangle&=&-i\int d^3 y\ \alpha^3(x-y)\frac{\delta}{\delta \Phi(y)}
\exp\left[i\int d^3 w\ d^3 z\ {\bar \alpha}^3(w-z)\Phi(w)\Pi(z)\right]\nonumber\\
&=&\int d^3 y\ d^3 w\ d^3 z\ \alpha^3(x-y){\bar \alpha}^3(w-z)\delta(w-y)\Pi(z)
\exp\left[i\int d^3 w\ d^3 z\ {\bar \alpha}^3(w-z)\Phi(w)\Pi(z)\right]\nonumber\\
&=&\int d^3 y\ d^3 z\ \alpha^3(x-y){\bar \alpha}^3(y-z)\Pi(z)
\exp\left[i\int d^3 w\ d^3 z\ {\bar \alpha}^3(w-z)\Phi(w)\Pi(z)\right]\nonumber\\
&=&\int\ d^3 z\ \delta(x-z)\Pi(z)\exp\left[i\int d^3 w\ d^3 z\ {\bar \alpha}^3(w-z)\Phi(w)\Pi(z)\right]\nonumber\\
&=&\Pi(x)\exp\left[i\int d^3 w\ d^3 z\ {\bar \alpha}^3(w-z)\Phi(w)\Pi(z)\right]\nonumber\\
&=&\Pi(x)|\Pi\rangle.
\label{eigenwert_equation_Pi}
\end{eqnarray}
The corresponding position eigenstate is given by

\begin{equation}
|\Phi[\Phi(x)]\rangle=\delta\left(\Phi[\Phi(x)]-\Phi^{\prime}[\Phi(x)]\right),
\label{eigenstate_Phi}
\end{equation}
where $\Phi(x)$ denotes the eigenvalue to the position eigenstate. The inner product for two states
referring to the field can be defined as

\begin{eqnarray}
\langle \Psi[\Phi(x)]|\Omega[\Phi(x)]\rangle=\int d \Phi(x)\ d^3 x\ \Psi[\Phi(x)]\ \Omega[\Phi(x)].
\end{eqnarray}
This implies that the inner product of a momentum eigenstate with a position eigenstate is given by

\begin{equation}
\langle \Phi|\Pi \rangle=\exp\left[i\int d^3 w\ d^3 z\ {\bar \alpha}^3(w-z)\Phi(w)\Pi(z)\right].
\label{innerproduct_PhiPi}
\end{equation}

\section{Derivation of the Propagator in Quantum Field Theory with Nonlocal Quantization Principle}

The general concept of the nonlocal quantization principle can now be studied in the context of a scalar
field theory. In this section the corresponding propagator is calculated in its general shape meaning that
no special assumption for the special choice of the quantization function is made so far. To determine the
propagator, it is necessary to formulate the generalized expression for the plane wave expansion of the
scalar field. The following expression for the quantum field

\begin{equation}
\hat \Phi(x,t)=\int \frac{d^3 p}{\sqrt{(2\pi)^3 2\omega_p}}\ \left(a(p)\int d^3 z\ \alpha^3(x-z) e^{i{\bf pz}-i\omega_p t}
+a^{\dagger}(p)e^{-i{\bf px}+i\omega_p t}\right)
\label{generalized_field}
\end{equation}
and its canoniocal conjugated variable

\begin{equation}
\hat \Pi(x,t)=\int \frac{d^3 p\ i\omega_p}{\sqrt{(2\pi)^3 2\omega_p}}\ \left(a(p)
\int d^3 z\ \alpha^3(x-z) e^{i{\bf pz}-\omega_p t}-a^{\dagger}(p)e^{-i{\bf px}+i\omega_p t}\right),
\end{equation}
where $a(p)$ and $a^{\dagger}(p)$ are the usual creation and annihilation operators
fulfilling the commutation relations 

\begin{equation}
\left[a(p),a^{\dagger}(p^{\prime})\right]=\delta(p-p^{\prime}),
\end{equation}
obey the generalized commutation relations given in ($\ref{generalization_commutator}$). This can be shown
as follows:

\begin{eqnarray}
\left[\hat \Phi(x,t),\hat \Pi(y,t)\right]&=&\frac{i}{(2\pi)^3}\int \frac{d^3 p}{\sqrt{2\omega_p}}
\ \int \frac{d^3 p^{\prime} \omega_{p^{\prime}}}{\sqrt{2\omega_p^{\prime}}} \int d^3 z
\left[\left[a(p),a^{\dagger}(p^{\prime})\right]\alpha^3(x-z)e^{i{\bf py}-i{\bf p^{\prime}z}}
+\left[a(p^{\prime}),a^{\dagger}(p)\right]\alpha^3(y-z)e^{i{\bf p^{\prime}x}-i{\bf pz}}\right]\nonumber\\
&=&\frac{i}{2(2\pi)^3}\int \frac{d^3 p}{\sqrt{\omega_p}}\ \int \frac{d^3 p^{\prime} \omega_p^{\prime}}
{\sqrt{\omega_p^{\prime}}} \int d^3 z \left[\delta^3(p-p^{\prime})\alpha^3(x-z)e^{i{\bf py}-i{\bf p^{\prime}z}}
+\delta^3(p^{\prime}-p)\alpha^3(y-z)e^{i{\bf p^{\prime}x}-i{\bf pz}}\right]\nonumber\\
&=&\frac{i}{2(2\pi)^3}\int d^3 p\ \int d^3 z \left[\alpha^3(x-z)e^{i{\bf py}-i{\bf pz}}
+\alpha^3(y-z)e^{i{\bf px}-i{\bf pz}}\right]\nonumber\\
&=&\frac{i}{2}\int d^3 z \left[\alpha^3(x-z)\delta^3(y-z)+\alpha^3(y-z)\delta^3(x-z)\right]\nonumber\\
&=&\frac{i}{2}\left[\alpha^3(x-y)+\alpha^3(y-x)\right]\nonumber\\
&=&i\alpha^3(x-y).
\end{eqnarray}
The corresponding propagator of the generalized quantum field ($\ref{generalized_field}$) is obtained by
inserting it to the genereal expression for the propagator in terms of the quantum field,
which is given by 

\begin{equation}
G(x-y)=\langle 0|T\left[\hat \Phi(x)\hat \Phi(y)\right]|0 \rangle,
\end{equation}
where $T$ denotes the time ordering operator $|0 \rangle$ denotes the vacuum state.
This leads to the following expression:

\begin{eqnarray}
G(x-y)&=&\int \frac{d^3 p}{(2\pi)^3 2p_0}\ \int d^4 z 
\left[\theta(x_0-y_0)\alpha^4(x-z)e^{-ipz+ipx}+\theta(y_0-x_0)\alpha^4(y-z)e^{-ipz+ipy}\right],
\end{eqnarray}
where the function $\alpha^4(x-y)$ has been defined as follows:

\begin{equation}
\alpha^4(x-y)=\alpha^3(x-y)\delta(x_0-y_0).
\end{equation}
By following the usual procedure of expresssing the $\theta$-function by an integral 

\begin{equation}
\theta(t)=\lim_{\epsilon \to 0} -\frac{1}{2\pi i}\int dE \frac{e^{-iEt}}{E+i\epsilon},
\end{equation}
the expression of the propagator can be transformed to

\begin{eqnarray}
G(x-y)=i\int \frac{d^3 p\ dE}{(2\pi)^4 2\omega_p}\frac{1}{E+i\epsilon}
\int d^4 z\ \left[\alpha^4(x-z)e^{i{\bf p}({\bf z}-{\bf y})-i(\omega_p+E)(z_0-y_0)}
+\alpha^4(y-z)e^{i{\bf p}({\bf z}-{\bf x})-i(\omega_p+E)(z_0-x_0)}\right].\nonumber\\
\end{eqnarray}
Changing of the variables according to, 

\begin{equation}
E^{\prime}=E+\omega_p,
\end{equation}
leads to

\begin{eqnarray}
G(x-y)&=&i\int \frac{d^3 p\ dE^{\prime}}{(2\pi)^4 2\omega_p}\frac{1}{E^{\prime}-\omega_p+i\epsilon}
\int d^4 z\ \left[\alpha^4(x-z) e^{i{\bf p}({\bf z}-{\bf y})-iE^{\prime}(z_0-y_0)}
+\alpha^4(y-z) e^{i{\bf p}({\bf z}-{\bf x})-iE^{\prime}(z_0-x_0)}\right]\nonumber\\
%&=&i\int \frac{d^3 p\ dE^{\prime}}{(2\pi)^4 2\omega_p}\frac{1}{E^{\prime}-\omega_p+i\epsilon}
%\int d^4 z\ d^4 z^{\prime}\left[\alpha^4(x-z)\delta^4(y-z^{\prime})+\alpha^4(y-z)\delta^4(x-z^{\prime})\right]
%e^{i{\bf p}({\bf z}-{\bf z^{\prime}})-iE^{\prime}(z_0-z^{\prime}_0)}
%\nonumber\\
&=&i\int \frac{d^3 p\ dE^{\prime}\ d^4 z}{2(2\pi)^4}\left[\frac{e^{i{\bf p}({\bf z}-{\bf y})
-iE^{\prime}(z_0-y_0)}}{\omega_p E^{\prime}-\omega^2_p+i\epsilon}\alpha^4(x-z)+x \longleftrightarrow y\right]\nonumber\\
&=&i\int \frac{d^3 p\ dE^{\prime}\ d^4 z}{2(2\pi)^4}\left[\frac{e^{i{\bf p}({\bf z}-{\bf y})-iE^{\prime}(z_0-y_0)}}
{\sqrt{{\bf p}^2+m^2}E^{\prime}-{\bf p}^2-m^2+i\epsilon}
\alpha^4(x-z)+x \longleftrightarrow y\right].
\end{eqnarray}
By setting

\begin{equation}
p_0=E^{\prime}
\end{equation}
one obtains the following generalized expression for the propagator in position space:

\begin{eqnarray}
G(x-y)&=&i\int \frac{d^3 p\ dp_0\ d^4 z}{2(2\pi)^4}
\left[\frac{e^{i{\bf p}({\bf z}-{\bf y})-ip_0 (z_0-y_0)}}{\sqrt{{\bf p}^2+m^2}p_0-{\bf p}^2-m^2+i\epsilon}
\alpha^4(x-z)+x \longleftrightarrow y\right]\nonumber\\
&=&i\int \frac{d^4 p\ d^4 z}{2(2\pi)^4}\left[\frac{e^{ip(z-y)}}{p^2-m^2+i\epsilon}
\alpha^4(x-z)+x \longleftrightarrow y\right].
\label{generalized_propagator}
\end{eqnarray}

\section{Special Model for the Nonlocal Quantization Principle}

The generalized nonlocal quantum field theory shall now be considered for a special choice of the quantization
function $\alpha^3(x-y)$ appearing in ($\ref{generalization_commutator}$). Since this function should be
symmetric in $x$ and $y$, should become smaller, if the distance becomes larger, should vanish at infinity
and should merge the delta function in the limiting case, a natural choice seems to be a Gaussian function,

\begin{equation}
\alpha^3(x-y)=\frac{e^{-\frac{1}{2}\frac{({\bf x}-{\bf y})^2}{\Delta}}}{\Delta\sqrt{2\pi}},
\label{Gaussian_function_3}
\end{equation}
where $\Delta$ describes the width of the Gaussian function. Accordingly $\alpha^4(x-y)$ reads

\begin{equation}
\alpha^4(x-y)=\delta(x_0-y_0)\frac{e^{-\frac{1}{2}\frac{({\bf x}-{\bf y})^2}{\Delta}}}{\Delta\sqrt{2\pi}}.
\label{Gaussian_function_4}
\end{equation}
If the expression for $\alpha^4(x-y)$ given in ($\ref{Gaussian_function_4}$) is inserted into
($\ref{generalized_propagator}$), one obtains the expression for the corresponding special
propagator

\begin{equation}
G(x-y)=i\int \frac{d^4 p\ d^4 z}{2(2\pi)^4}\left[\frac{e^{ip(z-y)}}{p^2-m^2+i\epsilon}\delta(x_0-z_0)
\frac{e^{-\frac{1}{2}\frac{({\bf x}-{\bf z})^2}{\Delta}}}{\Delta\sqrt{2\pi}}+x \longleftrightarrow y\right].
\end{equation}
By using the substitution

\begin{equation}
w=x-z,
\end{equation}
this expression can be transformed to

\begin{eqnarray}
G(x-y)&=&i\int \frac{d^4 p\ d^4 w}{2(2\pi)^4}\left[\frac{e^{ip(x-w-y)}}{p^2-m^2+i\epsilon}\delta(w_0)
\frac{e^{-\frac{1}{2}\frac{{\bf w}^2}{\Delta}}}{\Delta\sqrt{2\pi}}+x \longleftrightarrow y\right]\nonumber\\
&=&i\int \frac{d^4 p\ d^3 w\ dw_0}{2(2\pi)^4}\frac{1}{\Delta\sqrt{2\pi}}\left[\frac{e^{ip(x-y)}}{p^2-m^2+i\epsilon}\delta(w_0)
e^{-\frac{1}{2}\frac{{\bf w}^2}{\Delta}+i{\bf p}{\bf w}-ip_0 w_0}+x \longleftrightarrow y\right]\nonumber\\
&=&i\int \frac{d^4 p\ d^3 w}{2(2\pi)^4}\frac{1}{\Delta\sqrt{2\pi}}\left[\frac{e^{ip(x-y)}}{p^2-m^2+i\epsilon}
e^{-\frac{1}{2}\frac{{\bf w}^2}{\Delta}+i{\bf p}{\bf w}}+x \longleftrightarrow y\right].
\end{eqnarray}
By substituting {\bf w} through {\bf a} according to

\begin{equation}
{\bf a}=\frac{{\bf w}}{\sqrt{2\Delta}}-i\sqrt{\frac{\Delta}{2}}{\bf p},
\end{equation}
the corresponding integral can be solved and one obtains the final expression for the
propagator represented in position space of the quantum field theory with the nonlocal
quantization principle under consideration of the special choice ($\ref{Gaussian_function_3}$)
for the quantization function 

\begin{eqnarray}
G(x-y)&=&i\int \frac{d^4 p\ d^3 a}{(2\pi)^4}\frac{\Delta}{\sqrt{\pi}}
e^{-\frac{\Delta}{2}{\bf p}^2}\left[\frac{e^{ip(x-y)}}{p^2-m^2+i\epsilon}
e^{-{\bf a}^2}+x \longleftrightarrow y\right]\nonumber\\
&=&i\int \frac{d^4 p}{(2\pi)^4}\pi\Delta
e^{-\frac{\Delta}{2}{\bf p}^2}\left[\frac{e^{ip(x-y)}}{p^2-m^2+i\epsilon}
+x \longleftrightarrow y\right]\nonumber\\
&=&i\pi \Delta \int \frac{d^4 p}{(2\pi)^4} \left[\frac{e^{ip(x-y)}+e^{ip(y-x)}}
{p^2-m^2+i\epsilon}\right]e^{-\frac{\Delta}{2}{\bf p}^2}\nonumber\\
%&=&i2\pi \Delta \int \frac{d^4 p}{(2\pi)^4}\frac{e^{ip(x-y)}e^{-\frac{\Delta}{2}{\bf p}^2}}
%{p^2-m^2+i\epsilon}\nonumber\\
&=&i\Delta \int \frac{d^4 p}{(2\pi)^3}\frac{e^{ip(x-y)}
e^{-\frac{\Delta}{2}{\bf p}^2}}{p^2-m^2+i\epsilon}.
\end{eqnarray}
The corresponding expression for the propagator represented in momentum space is accordingly
given by

\begin{equation}
G(p)=\frac{i2\pi \Delta e^{-\frac{\Delta}{2}{\bf p}^2}}{p^2-m^2+i\epsilon}.
\end{equation}
This means that this quantum field theory in analogy to the situation in noncommutative geometry contains
no ultra violet divergences, since the propagator contains the damping factor $e^{-\frac{\Delta}{2}{\bf p}^2}$.
The special shape of the generalized propagator corresponds to the one calculated based on the coherent state
approach to noncommutative geometry \cite{Smailagic:2003rp},\cite{Smailagic:2003yb},\cite{Smailagic:2004yy}.
In \cite{Kober:2010um}, where the additional assumption of noncommuting momenta appears, has been found a
generalization of this propagator even containing no infra red divergences. Accordingly the presented generalized
quantum field theory, which is based on the concept of a nonlocal quantization principle represents at least
under certain consitions a promising candidate for an alternative theory to noncommutative geometry.

\section{Canonical Quantum Gravity with Nonlocal Quantization Principle}

If the nonlocal quantization principle is assumed to be a fundamental principle of nature, then of course
it has also be transferred to the quantum description of the gravitational field. In this section the
corresponding generalization of quantum geometrodynamics is considered. In quantum geometrodynamics
space-time is foliated into a time coordinate and a spacelike hypersurface. Accordingly the metric
field can be expressed in the following way:

\begin{equation}
g_{\mu\nu}=\left(\begin{matrix}N_a N^a-N^2 & N_b\\ N_c & h_{ab}\end{matrix}\right),
\end{equation}
where $h_{ab}$ denotes the three metric on the spacelike hypersurface, $N$ denotes the lapse function
and $N^a$ is the shift vector. The corresponding canonical conjugated momentum $\pi^{ab}$ reads 

\begin{equation}
\pi^{ab}=\frac{\partial \mathcal{L}}{\partial \dot h_{ab}}=\frac{\sqrt{h}}{16 \pi G}\left(K^{ab}-K h^{ab}\right),
\label{canonical_conjugated_momentum}
\end{equation}
with $K^{ab}$ describing the extrinsic curvature 

\begin{equation}
K_{ab}=\frac{1}{2N}\left(\dot h_{ab}-D_a N_b-D_b N_a\right).
\end{equation}
Since the generalization of the delta function should because of homogenity and isotropy of space just depend on
the distance of the two space points of the field operatopr and its canonical conjugated operator and with respect
to the canonical quantum description of general relativity the distance of two space points depends on the three
metric itself, the commutator depends on the three metric as well. This situation is similar to the generalized
quantization principle which has been developed in \cite{Kober:2011uj}. The distance $D$ between two points,
$x$ and $y$, on the spacelike hypersurface on which the three metric $h_{ab}$ is defined, is given by

\begin{equation}
D(x,y,h)=\int_x^{y} d\lambda\ \sqrt{h_{ab}(\lambda)\ \frac{dx^a}{d\lambda}\frac{dx^b}{d\lambda}}.
\end{equation}
In case of the variables of quantum geometrodynamics, this leads to the following generalized commutation
relation between the components of the three metric and its corresponding canonical conjugated quantity

\begin{equation}
\left[h_{ab}(x),p^{cd}(y)\right]=\frac{i}{2}\left(\delta_a^c \delta_b^d
-\delta_a^d \delta_b^c\right)\alpha^3 \left[D(x,y,h)\right].
\end{equation}
The corresponding nonlocal operators can be expressed as follows:

\begin{eqnarray}
\hat h_{ab}(x)|\Psi[h]\rangle=h_{ab}(x)|\Psi[h]\rangle,\quad
\hat \pi^{ab}(x)|\Psi[h]\rangle=-i\int d^3 y\ \alpha^3\left[D(x,y,h)\right]
\frac{\delta}{\delta h_{ab}(y)}|\Psi[h]\rangle.
\label{nonlocal_operators_gravity}
\end{eqnarray}
And the corresponding quantum constraints are obtained by inserting the operators ($\ref{nonlocal_operators_gravity}$)
to the corresponding classical constraints

\begin{eqnarray}
\hat {\mathcal{H}}(x)|\Psi[h]\rangle&=&\left[16\pi G G_{abcd}[\hat h(x)]\hat \pi^{ab}(x)\hat \pi^{cd}(x)
-\frac{\sqrt{\hat h(x)}}{16\pi G}\left(R[\hat h(x)]-2\Lambda \right)\right]|\Psi[h]\rangle=0,\nonumber\\
&=&\left[16\pi G G_{abcd}[h(x)]\int d^3 y\ d^3 z\ \alpha^3\left[D(x,y,h)\right]\alpha^3\left[D(x,z,h)\right]
\frac{\delta^2}{\delta h_{ab}(y)\delta h_{cd}(z)}
\right.\nonumber\\&&\left.
+\frac{\sqrt{h(x)}}{16\pi G}\left(R[h(x)]-2\Lambda \right)\right]|\Psi[h]\rangle=0,\nonumber\\
\hat {\mathcal{H}}_a(x)|\Psi[h]\rangle&=&-2 D_b \hat \pi_a^b(x)|\Psi[h]\rangle=
-2 D_b h_{ac}(x)\int d^3 z\ \alpha^3 \left[D(x,y,h)\right]\frac{\delta}{\delta h_{bc}(y)}|\Psi[h]\rangle=0,
\end{eqnarray}
where $G_{abcd}$ is defined according to

\begin{equation}
G_{abcd}=\frac{1}{2\sqrt{h}}\left(h_{ac}h_{bd}+h_{ad}h_{bc}-h_{ab}h_{cd}\right).
\end{equation}
If the special case of the assumption of a Gaussian function for the quantization function is considered,
this function is in case of quantum geometrodynamics of the following shape:

\begin{equation}
\alpha^3[D(x,y,h)]=\frac{e^{-\frac{1}{2}\frac{[D(x,y)]^2}{\Delta}}}{\Delta \sqrt{2\pi}},
\end{equation}
where the squared difference between the space-time points has to be replaced by the distance
function $D(x,y,h)$ depending on the three metric $h_{ab}$, too.

\section{Path Integral Quantization}

Of course it has to be possible to formulate the generalized quantization principle, which is given in
($\ref{generalization_commutator}$) by generalization of the usual commutation relations belonging
to canonical quantization, within the path integral formalism as well. To obtain such a formulation in
the framework of path integrals, first the general expression of the transition amplitude between
two states separated by an infinitesimal time interval has to be modified according to the
generalization of the canonical quantization condition. The transition amplitude between a state
$|\Phi(t)\rangle$ at time $t$ and a state $|\Phi^{\prime}(t+dt)\rangle$ at time $t+dt$ is given by

\begin{eqnarray}
\langle \Phi^{\prime}(t+dt)|\Phi(t)\rangle
&=&\int d \Pi\ \langle \Phi^{\prime}|\exp\left[-i\ dt\int d^3 x\ \hat {\mathcal{H}}\left(\hat \Phi(x),\hat \Pi(x)\right)\right]
|\Pi \rangle \langle \Pi|\Phi\rangle\nonumber\\
&=&\int d \Pi\ \langle \Phi^{\prime}|\exp\left[-i\ dt\int d^3 x\ {\mathcal{H}}\left(\Phi(x),\Pi(x)\right)\right]
|\Pi \rangle \langle \Pi|\Phi\rangle.
\label{transition_amplitude_A}
\end{eqnarray}
Usually the inner product between an arbitrary eigenstate of the field operator and an arbitrary eigenstate of the
momentum operator reads $\rangle \langle \Pi|\Phi\rangle=\exp\left[i\int d^3 x\ \Phi(x)\Pi(x)\right]$. In case of
the nonlocal quantization principle the general inner product is given by ($\ref{innerproduct_PhiPi}$). Inserting
of ($\ref{innerproduct_PhiPi}$) to ($\ref{transition_amplitude_A}$) leads to

\begin{eqnarray}
\langle \Phi^{\prime}(t+dt)|\Phi(t)\rangle&=&\int d \Pi\ \exp\left[-i\ dt\int d^3 x\ {\mathcal{H}}\left(\Phi(x),\Pi(x)\right)
+i\int d^3 w\ d^3 z\ {\bar \alpha}^3(w-z)\left(\Phi^{\prime}(w)-\Phi(w))\right)\Pi(z)\right].\nonumber\\
\label{transition_amplitude_B}
\end{eqnarray}
By using the expression of the transition amplitude for an infinitesimal time interval in case of the nonlocal quantization
principle, the corresponding transition amplitude for a non infinitesimal time interval can be determined as usual in
the following way:

\begin{eqnarray}
\langle \Phi(t_b)|\Phi(t_a)\rangle&=&\lim_{N \to \infty}\int d \Phi_1...d \Phi_N
\langle \Phi(t_b)|\Phi_N \rangle...\langle \Phi_1 |\Phi(t_a)\rangle\nonumber\\
&=&\lim_{N \to \infty}\int \prod_{k=1}^N d \Phi_k \prod_{k=0}^N d \Pi_k
%\nonumber\\&&\quad\quad\quad\quad
\exp\left[i\sum_{k=1}^{N+1}\int d^3 x\left(\int d^3 z\ \bar \alpha^3(x-z)\left(\Phi_k(x)-\Phi_{k-1}(x)\right)\Pi_{k-1}(z)
\right.\right.\nonumber\\&&\left.\left.
\quad\quad\quad\quad\quad\quad\quad\quad\quad\quad\quad\quad\quad\quad\quad\quad
\quad\quad\quad\quad\quad\quad\quad\quad\quad\quad\quad\quad
-\mathcal{H}\left(\Phi_k(x),\Pi_{k-1}(x)\right)\right)\right]\nonumber\\
&=&\int \mathcal{D}\left[\Phi(x)\right]\mathcal{D}\left[\Pi(x)\right]\exp \left[i\int_{t_a}^{t_b} dt \int d^3 x
\left(\int d^3 z\ \bar \alpha^3(x-z)\Pi(x)\dot \Phi(z)-\mathcal{H}\left(\Phi(x),\Pi(x)\right)\right)\right].
\end{eqnarray}
This represents the analogue generalization from the usual quantization principle to the nonlocal
quantizastion principle in the path integral formalism.

\section{Summary and Discussion}

In this paper has been suggested a generalization of the concept of quantization within quantum field theory.
The generalization consists in the assumption that the commutator between the operator describing the
field, which is quantized, and the operator describing its corresponding canonical conjugated momentum
does not vanish, if the two operators refer to different space-time points, as it is presupposed in the
usual setting of quantum field theory. Accordingly the delta function appearing on the right hand side
of the quantization condition usually, has to be replaced by another function, which does not vanish,
if the space-time points the two operators within the commutator refer to are not equal. Because of
the postulate of homogenity and isotropy of space, this function should be assumed just to depend on
the distance between the two space coordinates. According to this nonlocal quantization principle
generalized field operators have to be defined fulfilling this generalized quantization conditions.
Based on the generalized expression of a scalar field operator, the corresponding generalized expression
for the propagator has been calculated, which of course directly depends of the generalization of the
delta function appearing in the generalized quantization condition. If the natural assumption is made
that the quantization function is described by a Gaussian function, then one obtains a special shape
for the propagator with an exponential containg the negative square of the three momentum as damping
factor. Thus the propagator corresponds to the propagator found in the coherent state approach
of noncommutative geometry and contains no ultra violet divegences. This is a very important hint that
this generalization of quantum field theory omitting the postulate of locality with respect to the
basic quantization condition could indeed represent a promising candidate for a generalization of
quantum field theory similar to quantum field theory on noncommutative space-time. Since such a
quantization principle should be interpreted as fundamental property of nature, it has analogously
to be transferred to the quantum description of all field theories including general relativity.
This implies the necessity to generalize the canonical quantum description of general relativity
as well. In this paper has been given a consideration with respect to quantum geometrodynamics, which
has not been elaborated completely yet. But the quantization function has in each case to be assumed
to depend on the gravitational field, since the distance between to space points depends on the
gravitational field. This leads to a similarity to a generalized quantization principle within
canonical quantum gravity \cite{Kober:2011uj} in accordance with a generalized uncertainty in quantum
mechanics, where the right hand side of the quantization condition depends on the operators themselves.
It has also to be possible to incorporate the nonlocal generalization of the quantization principle to
path integrals. This has been done by replacing the usual inner product between an arbitrary eigenstate
of the field operator and an arbitrary eigenstate of its canonical conjugated variable by the
corresponding generalized one. Based on this generalization of path integrals it is in principle
possible to incorporate the nonlocal quantization principle to gravity by using the path integral
formalism. In general it could be very interesting with respect to further research projects, to
formulate a full quantum description of general relativity based on this concept. Perhaps because of
the ultraviolet behaviour of the commutator it can even contribute to circumvent the problem of
nonrenormalizability of the quantrum description of gravity in the framework of usual quantum field
theory. An additional important task would be to determine an upper limit for the parameter $\Delta$
within the Gaussian function, since the generalization of the quantization principle must not get in
conflict with the usual description of quantum field theory at low energies.

\end{document}